\documentclass[twocolumn,prl,aps,showpacs,a4paper,floatfix]{revtex4}

\usepackage{graphicx}
\usepackage{epsfig}
\usepackage{psfig}

\begin{document} 
\psdraft 
\psfull

\title{Dynamic  Phase Transitions in Cell Spreading}
\author{Hans-G\"{u}nther D\"{o}bereiner, Benjamin Dubin-Thaler, Gr\'{e}gory Giannone, 
Harry S. Xenias, and Michael P. Sheetz 
}
\affiliation{
Department of Biological Sciences, Columbia University, New York, NY 10027
}

\email{hgd@biology.columbia.edu}

\begin{abstract}
We monitored isotropic spreading of mouse embryonic fibroblasts on fibronectin-coated substrates. 
Cell adhesion area versus time was measured via total internal reflection fluorescence microscopy.   Spreading proceeds in well-defined phases. 
We found a power-law area growth 
with distinct exponents $a_i$ in three sequential phases, which we denote basal $(a_1=0.4\pm0.2)$, continous $(a_2=1.6\pm0.9)$ and contractile $(a_3=0.3\pm0.2)$ spreading.   High resolution differential interference contrast  microscopy
was used to characterize local membrane dynamics at the spreading front. Fourier power spectra of membrane velocity reveal the sudden development of
periodic  membrane retractions at the transition from continous to contractile spreading. We propose that the classification of cell spreading into phases
with distinct functional characteristics and protein activity patterns serves as a paradigm for a general program of a phase classification of cellular phenotype. 
Biological variability is drastically reduced when only the corresponding phases are used for comparison across species/different cell lines. 
\end{abstract}

\maketitle

Cells need to be mobile in order to perform many critical biological functions.
The reorganization of  extracellular matrix in wound healing, the positioning of nerve cells, or  the engulfment  of  bacteria in the immune reaction of white blood cells are particular
examples \cite{GeneralRef}.  Accomplishing this variety of functions requires a diverse set of mechanisms and proteins. 
Most components of the molecular machinery of actin-based motility have been identified \cite{Stossel93,Carlier01,Polard03}.   It has been possible to perform experiments with reconstituted systems of Listeria propulsion \cite{Bernheim02,Giardini03} for which detailed elastic models have been developed \cite{Gerbal00}.   In vivo, cell spreading on matrix-coated surfaces provides a simplified system of analyzing motile behavior.  A substantial amount of experimental  and theoretical work  has been done along these lines \cite{Sheetz,Geiger,Dembo,Schwarz}.  However, only quite recently, quantitative experiments of whole cell spreading
and subsequent migration could be performed with high spatial and temporal resolution  \cite{Dubin2004,Giannone2004}.  We found that there are well defined and distinct  states of spreading. It is the goal of this work to show that these states can indeed be considered phases of motility by demonstrating the existence of dynamic phase transitions between them.  

\begin{figure}[h]
\begin{center}
 \leavevmode
\epsfig{file=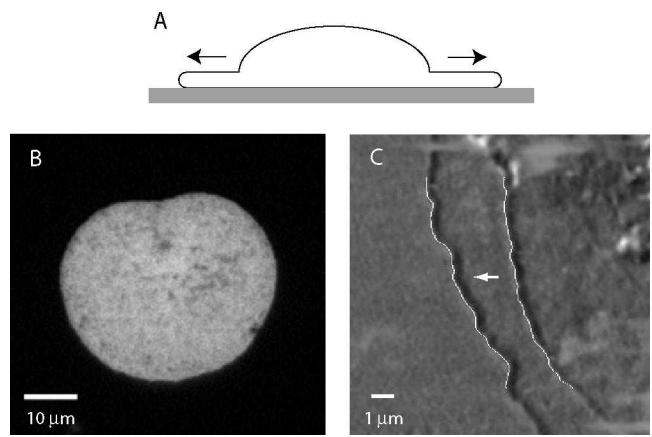,width=8.5cm}
 \caption{
 A:  During cell spreading, a thin lamellipodial sheet extends from the cell body onto the substrate.  B: Total internal reflection fluorescence micrograph of a spreading cell. The bright region corresponds to the area adhered to the substrate.
   C: Two overlayed snapshots of the leading membrane edge of a lamellipodium moving from right to left are shown in differential interference contrast.  The edge position is marked with a white contour overlay.  }
   \label{fig:Lamellipodium}
 \end{center} 
\end{figure}

Spreading  cells  extend a 200 nm thick 
sheet  called the lamellipodium from the cell body onto the substrate, see Fig.~\ref{fig:Lamellipodium}.
This process is driven by actin polymerization at the leading membrane edge, the precise mechanism of which  is still under debate \cite{Dickinson02, Mogilner03}.  The meshwork of actin fibers is crosslinked by various proteins. The molecular motor myosin II enables the meshwork to contract
by moving along actin fibers and relative to other cytoskeletal elements. Thus,  the lammelipodium is an active gel enclosed in a flat membrane bag adhering to the substrate.
The physics of active gels has recently attracted a lot of attention. Rheological experiments of simple mixtures of purified actin and myosin solutions \cite{Humphrey02} and  quite general  theoretical modeling \cite{Kruse03,Kruse04}  have been carried out.  There are dynamic phase transitions involving extended and contracted actin density states  as a function of myosin-actin coupling strength \cite{Kruse03}.  
We will show that our cellular system exhibits similar transitions which express themselves prominently in the dynamics of the leading membrane edge.   

Mouse embryonic fibroblasts (MEF) were allowed to settle onto fibronectin coated glass slides and observed with
either total internal reflection fluorescence (TIRF) or differential interference contrast (DIC) microscopy. 
Fibronectin links the cell membrane to the extracellular matrix proteins mimicking natural cell adhesion.
 TIRF studies were performed at a moderate spatial and temporal resolution  to capture overall spreading characteristics of the whole cell. Multiple cells could be studied simultaneously. High resolution DIC was used to characterize local membrane dynamics.  Details of the methods may be found in our earlier work \cite{Giannone2004,Dubin2004}. 

\begin{figure}[h]
\begin{center}
 \leavevmode
\epsfig{file=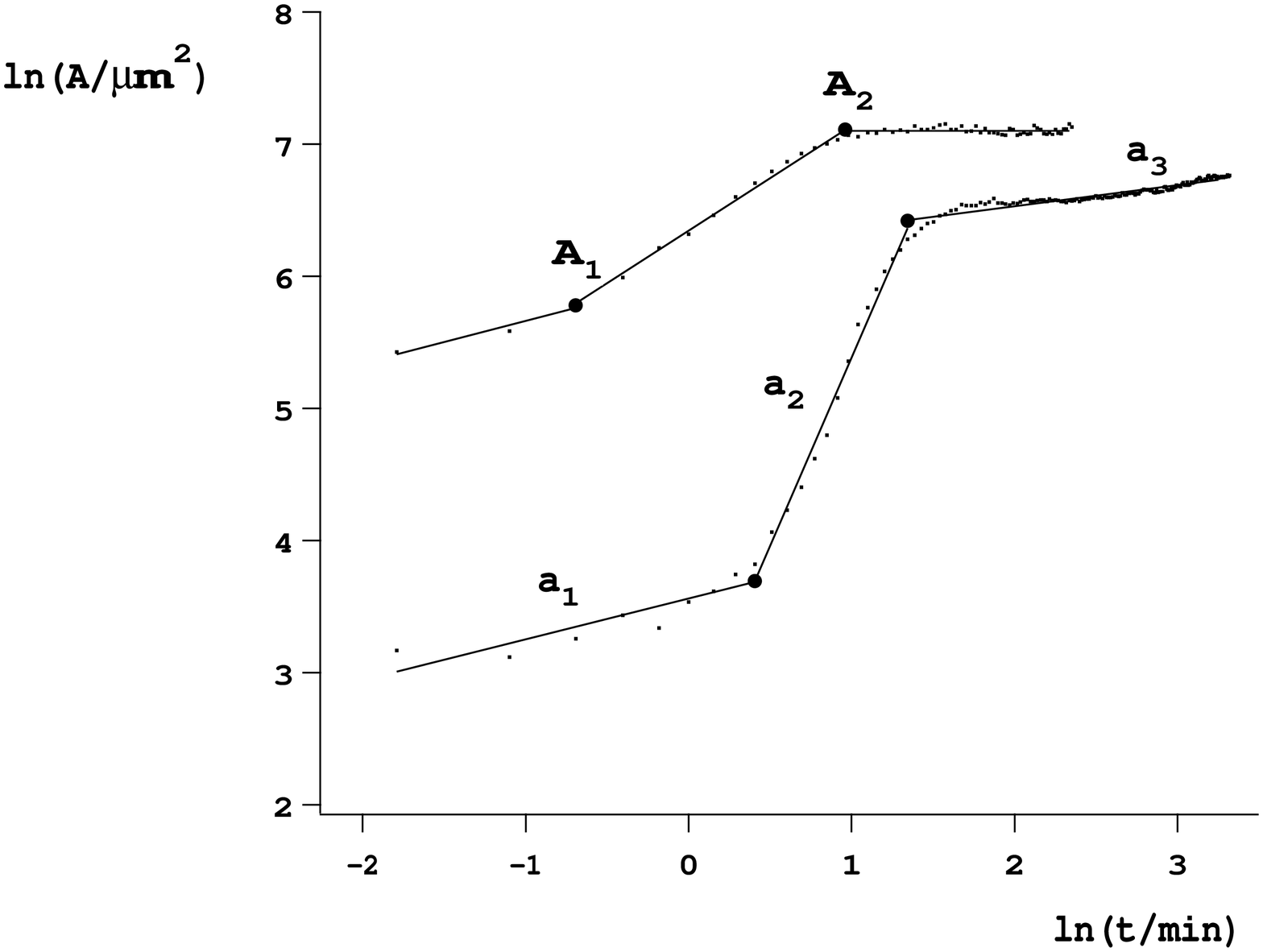,width=9cm}
 \caption{Adhesion area in isotropically spreading fibroblasts grows with a power law in time.  Different but constant 
 exponents $a_i$ in the various phases of spreading are evident in a double logarithmic plot. Exponents have been determined by fitting a piecewise linear function to the data, see Fig.~\ref{fig:Exponents}. Adhesion areas $A_i$ at the  transition points are indicated. }
  \label{fig:Scaling}
 \end{center} 
\end{figure}
\begin{figure}[h]
\begin{center}
\leavevmode
\epsfig{file=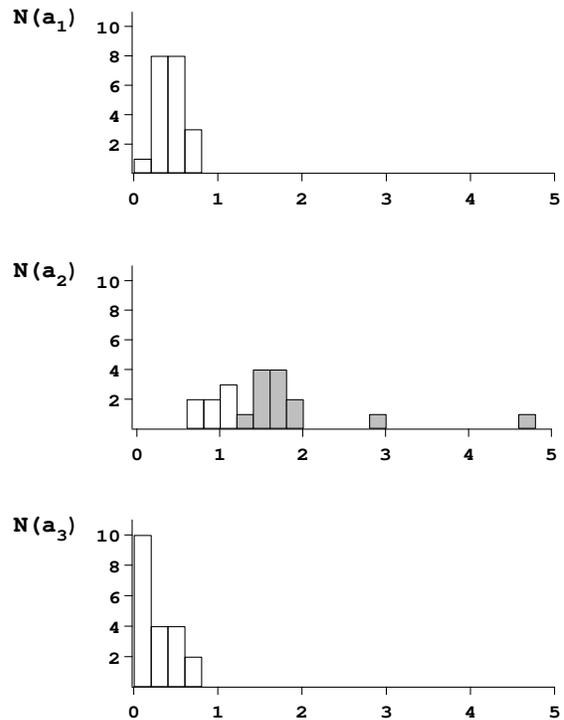,width=8cm}
\caption{Histograms of area growth exponents as obtained from the slopes of double logarithmic plots of
adhesion area versus time, like the ones shown in Fig.~\ref{fig:Scaling}.  We have analyzed 20 cells in total. The middle (continous spreading) phase exhibits clustering corresponding to small (open bars) and large 
(filled bars) area growth during that phase.  }
\label{fig:Exponents}
\end{center} 
\end{figure}

Membrane adhesion area $A$ during spreading was best monitored using TIRF.  Distinct classes of angular isotropic and anisotropic spreading cells were found \cite{Dubin2004}. 
In the following, we limit ourselves to the isotropic class lacking filopodia. 
Close inspection of double logarithmic plots of adhesion area $A$ over time reveals three phases with distinctly different power law growth, as seen in Fig.~\ref{fig:Scaling}. 
We define area growth exponents $a_i$  via 
\begin{equation}
A(t) \sim  t^{a_i}\ ,
\end{equation}
where $i$ denotes the subsequent phases.
Initially, there is a basal phase where cells test the suitability of the substrate to adhere and area growth is mimimal. We find $a_1 = 0.4 \pm 0.2$. Then
follows a  phase  of fast continuous spreading, which is characterized by $a_2 = 1.6 \pm 0.9$. Finally, the cell slows down again exhibiting a sub-linear area growth\
with $a_3 = 0.3 \pm 0.2$. We will see below that the latter phase is characterized by periodic local contractions of the cell \cite{Giannone2004}. Nevertheless, the mean  area growth leads to an effective power law behavior
also  in this phase.  Histograms  of exponents $a_i$ for the three phases are shown in Fig.~\ref{fig:Exponents}. There is a clear distinction of fast area growth in the continuous spreading phase with
a rather broad  distribution of the exponent $a_2$. However, we find two narrow clusters when discriminating with respect to the relative area growth, $A_2 / A_1$, during that phase,  where $A_i$ denotes the adhesion area at the transition from phase $i$ to $i+1$.  Small ($A_2/A_1 < 5$) or  large ($A_2/A_1 > 5$) area increases correspond to small ($m_2 = 0.9 \pm 0.2 $) or large ($m_2 = 1.6 \pm 0.2 $) exponents, respectively. In addition, there were two single cells with even larger exponents $m_2$  which we excluded from the cluster average. 

The transition from continuous to contractile spreading was further monitored using high resolution DIC.  
A suitable isotropically  spreading cell was chosen and a well-resolvable and approximately straight membrane segment  was selected for prolonged observation, see Fig.\ref{fig:Lamellipodium}~B. Time-lapse sequences were obtained at video rate.  Movies were digitized at $1/\Delta t = 3 Hz$.  Individual frames are counted using an index $n$. The cell edge is determined with a custom C program by a local contour algorithm \cite{Doeb97} allowing nanometer accuracy. We obtain a sub-pixel resolution of 15 nm for relative displacements, which translates into a minimal detectable velocity of 45 nm/s between frames.  Further analysis proceeds using a cartesian coordinate system  where the average membrane orientation is taken as the fixed  y-axis. Points on the membrane are then labeled by  their y-coordinates $y_j$ and the membrane velocity $v_j(n)=\Delta x_j(n)/\Delta t$ is measured along the x-axis which is normal to the average membrane orientation. 
\begin{figure}[h]
\begin{center}
 \leavevmode
  \epsfig{file=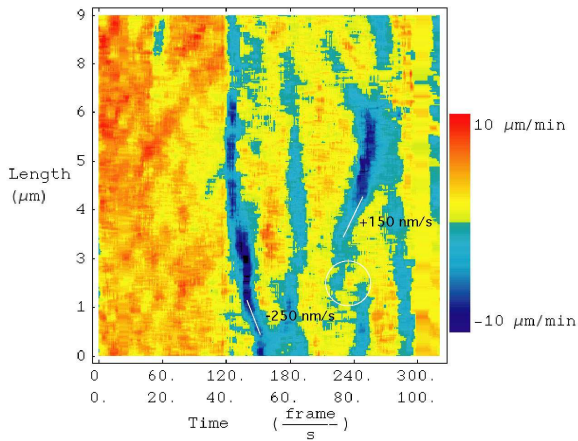,width=8cm} 
 \caption{Normal velocity map  of the particular membrane segment marked in Fig.~\ref{fig:Lamellipodium}B as a function of time. Note the two qualitatively different sections before and after time $t=40\ s$ corresponding to a continous and a periodically contractile spreading phase, respectively. The period of  the latter is $T = 17 \pm 4\ s$. 
The speeds of lateral waves of maximum contraction velocity are indicated. The encircled region
marks a phase shift of a contraction.}
  \label{fig:Velocity}
 \end{center} 
\end{figure}

\begin{figure}[h]
\begin{center}
 \leavevmode
\epsfig{file=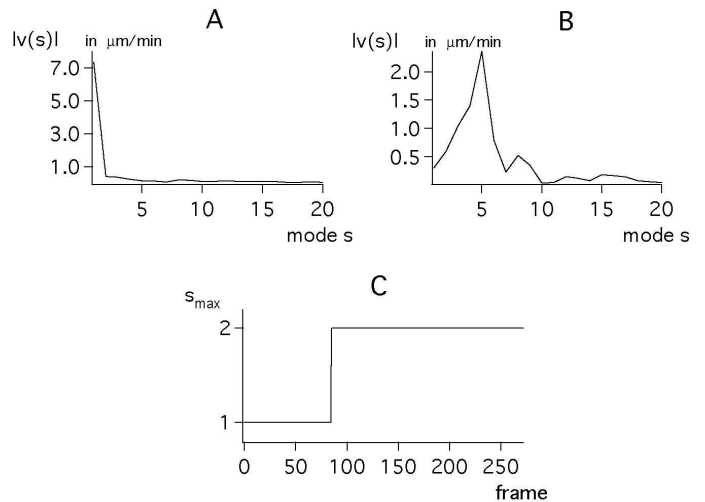,width=9cm}
\caption{Fourier spectrum (see Eq.~\ref{spectrum}) of the velocity  map in Fig.~\ref{fig:Velocity} for the two  different spreading phases below (A,$N=120$) and above (B,$N=200$) frame number $n=120$. The spectrum is spatially averaged over $70$ points between position $2.0\ \mu m$ and $3.8\ \mu m$ along the contour.   
  The transition between continous and contractile spreading  is characterized by a sharp shift in the position of the maximum of the Fourier spectrum.
 The boundary maximum in panel~A corresponds to a mean velocity of $7.4\ \mu m$ in the continous spreading phase.
 The peak at $s_{\rm max}=5$ in panel~B corresponds to a repeat time $T=17\pm 4\ s$ for the contractile 
 spreading phase, see Eq.~\ref{period}. The mean velocity $0.3\ \mu m$ is small. 
  The lower panel~C depicts the  peak position of the spectrum taken in a running  time window with a width of $N=50$ frames, corresponding to the repeat time $T$, as a function of  the first frame number of the window. Note that the peak position depends on the width of the window used for Fourier transformation.}
  \label{fig:Transition}
 \end{center} 
\end{figure}

A typical velocity map along the contour over time is shown in Fig.~\ref{fig:Velocity}. 
We find that a region of continous, uninterrupted spreading (red shadows) precedes a sequence of periodic membrane retraction events (blue stripes).  
These two different states of membrane dynamics correspond to the continuous spreading and contractile phase of the lamellipodium, found above. 
The two phases can be clearly  distinguished using
the discrete Fourier transformation $v(s)$ of the velocity map  $v(n)$ defined as
\begin{equation}
v_j(s) = \frac{1}{N}\sum_{n=1}^{N} v_j(n)\ \exp{\left(2\pi i\frac{(n-1)(s-1)}{N}\right)},
\label{spectrum}
\end{equation}
where N is the number of frames. Averages are taken over spatial regions of interest. 
	The continuous spreading phase is characterized by a strong boundary maximum of the spectrum $|v_j(s)|$ at 
\mbox{$s=1$}, see Fig.~\ref{fig:Transition}~A.   In contrast, in the contractile phase the spectrum develops a 
pronounced peak at $s=s_{\rm max}$, see Fig.~\ref{fig:Transition}~B, which signals oscillatory behavior with a period
\begin{equation}
T = \Delta t \frac{N}{s_{\rm max}-1}\ .
\label{period}
\end{equation}
Thus, the peak position of the power spectrum  serves as an excellent phase indicator.
We calculate the spectrum inside a small time window 
-  with a width on the order of the repeat time - and sweep across the phase boundary. 
Indeed, there is a well defined transition between the two phases as seen in Fig.~\ref{fig:Transition}~C. 
However, 	the periodic contractions do not take place simultaneously along the leading edge, 
see Fig.~\ref{fig:Velocity}. In fact, there are  lateral waves of maximum contraction velocity running in both directions.  These waves have a speed on the order of  $200\ nm/s$.
 Moreover, there are sudden phase shifts of the periodic contractions up to half a period, see encircled region in Fig.~\ref{fig:Velocity}.

To summarize our experimental findings, we have seen clear signatures of dynamic phase transitions
in the spreading behavior of  MEF  cells. Since actin polymerization does not stop during membrane retraction events  \cite{Giannone2004}, one concludes that the actin network  contracts
and/or is actively pulled back by myosin motor activity.  Kruse et al. \cite{Kruse03} have modeled oriented
fibers connected into a network by molecular motors. They find an instability of homogeneous
fiber density towards a contracted state  as a function of fiber-motor coupling strength. Moreover, their generic theoretical model allows for oscillatory solutions.   Our cellular system exhibits similar behavior. Indeed, the periodic contractions are absent when myosin light chain kinase (MLCK), which activates myosin, is inhibited \cite{Giannone2004}. 

In the following, we give a systems biology oriented view of cell motility. 
Several questions arise: What is the functional role of the phases described above?
How are these dynamic phases of the structural motility network regulated and connected to the signaling
network?  Can we disentangle the complex set of motility related proteins  and simplify description  by considering functional modules and hierarchical levels of control? 

During the initial spreading phase, a MEF cell assembles the cytoskeletal structure necessary to probe the mechanical suitability of the substrate in the following contractile phase where it  periodically pulls on the substrate. Indeed, cells require stiff substrates for growth and move toward stiffer regions \cite{Schwarz}. The machinery for
this stiffness sensing is organized into i) the basic structural elements consisting of the actin cytoskeleton, the myosin II motors, as well as the plasma membrane,  ii)  factors directly regulating protein coupling
strength and activity, and,  finally,   iii) these regulatory proteins are  controlled by   a signaling network coordinating spatially distant and/or  logically separate functional events in the cell. 
	In order to link these cellular components to the physics of dynamic phase transitions, we note that
 it is the basic structural elements which exhibit the various phases we have identified in this work.
The phase parameters are given by the regulating proteins, e.g., MLCK activity and concentration. The different dynamic phases correspond
 to  functional regions in the regulating para\-
 meter space. The trajectories in this parameter space are determined
 by the cellular signaling network.  This hierarchical identification provides an immediate conceptual advantage: 
The topology of the motility phase diagram is independent of the complex signaling network, i.e., the relative
positions of all the motility phases do not depend on the trajectories in parameter space followed by the cell. 
Indeed, the basic phase characteristics can be probed and modeled separately.  This was demonstrated using the observed linear relationship
between the period of  contractions and the lamellipodia width \cite{Giannone2004}.
  We find that  the contraction period is the same for equal  lamellipodia width, 
independent from the variations in the biochemical pathway(s) induced in order to achieve a certain width. 
On the other hand, cell motility is a unique case where the interconnections of the structural and signaling networks, which depend on gene expression,
can be probed in order to establish a quantitative link between phenotype and genotype. We are currently identifying functional modules
in a large scale screening of spreading phenotypes across various fibroblast cell lines with mutant genotypes.

The idea of phases in cell behavior can be applied  quite generally. Phases of motility should be considered
analogous to the phases of the cell cycle, phases of varying metabolic activity or different protein expression. 
We propose to classify cellular behavior  in well defined  phases.  Their number
will be considerable less than an enumeration of concentration and activity levels of all molecular components of the cell. Thus, one can hope to accomplish a simplified description. 
  Currently, phase classification is not generally done and cellular phenotype cannot be sensibly compared across different genotypes. We expect that  some fraction of  the variability  encountered in biological experiments and the often conflicting results between laboratories  stem from the fact that findings corresponding to different cellular phases and boundary conditions are spuriously compared to each other. 
	In conclusion, we feel that the classification of motility in phases can serve as a paradigmatic example for
a powerful general ordering principle in quantitative biology.

\vspace*{1cm}
\begin{acknowledgments}
This work was funded by the Deutsche Forschungsgemeinschaft  and the National Institutes of Health via grant GM 036277.  HGD would like to thank Adam Meshel  and Ana Kostic for helpful discussions.
\end{acknowledgments}


\end{document}